# Multi-Spectral Near Perfect Metamaterial Absorbers Using Spatially Multiplexed Plasmon Resonance Metal Square Structures

**Boyang Zhang,[1] Joshua Hendrickson,[2] and Junpeng Guo[1,*]**

[1]*Department of Electrical and Computer Engineering, University of Alabama in Huntsville, Huntsville, Alabama 35899, USA*
[2]*Sensors Directorate, Air Force Research Laboratory, Wright Patterson Air Force Base, Ohio 45433, USA*
*Corresponding author: guoj@uah.edu*

Near perfect infra-red light absorption at multiple wavelengths has been experimentally demonstrated by using multiplexed metal square plasmonic resonance structures. Optical power absorption over 95% has been observed in dual-band metamaterial absorbers at two separate wavelengths and optical power absorption over 92.5% has been observed in triple-band metamaterial absorbers at three separate wavelengths. The peak absorption wavelengths are primarily determined by the sizes of the metal squares in the multiplexed structures. Electrical field distributions in the middle of the dielectric spacer layer were calculated at the peak absorption wavelengths. It is found that the strong light absorption corresponds to local quadrupole plasmon resonance modes in the metamaterial structures.

## 1. Introduction

Since the observation of anomalous light absorption in metal gratings by R. W. Wood [1] in 1902, strong light absorption in a variety of metal structures has been extensively investigated [2-8]. In the past few years, perfect absorption in metamaterials has been investigated in the gigahertz [9-10], terahertz [11-12], and optical frequency regimes [13-19]. It is understood that perfect absorption in metamaterials is due to electromagnetic resonances in the structures and that the number of absorption bands depends on the number of resonant modes in the structures. Dual and triple spectral-band metamaterial perfect absorbers have been investigated by using various plasmon resonance metal structures in the infrared [20-26], THz [27-31] and microwave [32-36] frequency ranges. Recently, wideband metamaterial perfect absorbers made by multiplexing two different sized metal squares in the unit cells were demonstrated in the THz and infra-red regimes [37, 38]. In this paper, we present numerical designs and the experimental demonstration of dual and triple spectral-band near perfect metamaterial absorbers using spatially multiplexed metal square plasmonic resonance structures in the mid-wave infrared regime.

## 2. Design of Multi-spectral Metamaterial Perfect Light Absorbers

**A. Multiplexed structures for perfect light absorption**

Figures 1 (a) and (b) show the dual and triple spectral-band metamaterial absorber structures, respectively. Both structures consist of a thin film gold square array and a thick gold film separated by a dielectric spacer layer made of IR transparent magnesium fluoride ($MgF_2$). The thick gold film sits on a silicon substrate and is thick enough that no transmission can occur when light is incident from above. Absorptance can, therefore, be calculated from the reflectivity of the sample surface. In the dual spectral-band absorbers, two thin film gold squares with different sizes are multiplexed diagonally in the unit cell. For the triple spectral-band absorbers, four thin film gold squares of three different sizes are multiplexed in the unit cell. Among the four gold squares, the two smallest squares have the same size and are placed diagonally from one another. The structures of both the dual-band and triple-band metamaterial absorbers are laterally symmetric to ensure polarization independence for normal incidence radiation.

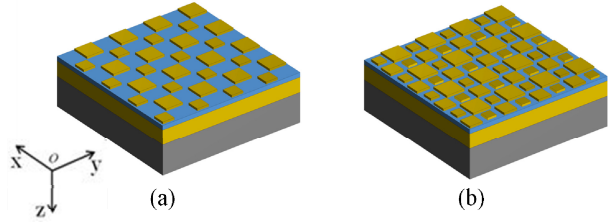

Fig. 1. (a) The dual spectral-band perfect light absorber structure and (b) the triple spectral-band perfect light absorber structure.

**B. Design of dual-spectral band metamaterial perfect absorbers**

A finite-difference time-domain (FDTD) software code developed by Lumerical Solution, Inc. was used to simulate the dual-band and triple-band perfect light metamaterial absorbers. In the simulations, the Lorentz-Drude material model is used for calculating the optical constants of gold films [39]. The electric permittivity of the $MgF_2$ layer is obtained from reference [40]. Periodic boundary conditions are used in the x and y dimensions to accommodate the periodicity of the structures. Perfectly matched layers (PMLs) are used above and below the absorber structure in the direction of propagation. The thick gold metal film is set at 250 nm for eliminating power transmission to the substrate. The incident light propagation is normal to the surface in the z direction and the polarization is in the x direction. The amplitude of the incident electric field is 1.0 V/m. A two-dimensional (2D) optical power monitor is placed above the absorber surface at a location between the source and the upper PML boundary to obtain the optical power reflectivity. A 2D electric field monitor is placed in the middle of the $MgF_2$ spacer layer to obtain the field distributions inside the gap between the gold square array and the thick gold film.

Figure 2 shows the calculated reflectivity from dual-band absorbers with two different size metal squares in the unit cell and for two different film thicknesses. In each case the period of the unit cell is 2.4 μm in both the x and y directions and the thickness of the $MgF_2$ spacer layer is 75 nm. Fig. 2 (a) shows the optical reflectivity from the dual-band absorber with 0.90 μm and 1.30 μm gold squares of 77 nm thickness in the unit cell. The absorption reaches a peak value of 99.4% at 3.49 μm wavelength and 98.8% at 4.87 μm wavelength, respectively. In Fig. 2(b) the film thickness is changed to 100 nm and the absorption now reaches 99.4% at 3.53 μm wavelength and 94.3% at 4.94 μm wavelength. Larger gold squares of 0.95 μm and 1.35 μm are used in Fig. 2(c) with the film thickness set back to 77 nm. Peak absorptions of 99.3% at 3.67 μm and 96.9% at 5.01 μm now occur. Finally in Fig. 2 (d), the thickness of the larger sized gold squares is increased to 100 nm. This results in peak absorptions of 99.5% at 3.72 μm and 93.4% at 5.166 μm. From the simulations it can be seen that, among the four designs, the dual-spectral band absorber with 0.90 μm and 1.30 μm gold squares of 77 nm thickness gives the best absorption performance at two separate wavelengths. The small resonance dip at 2.4 μm wavelength is due to the coupling of in-plane diffraction waves in the metal square array.

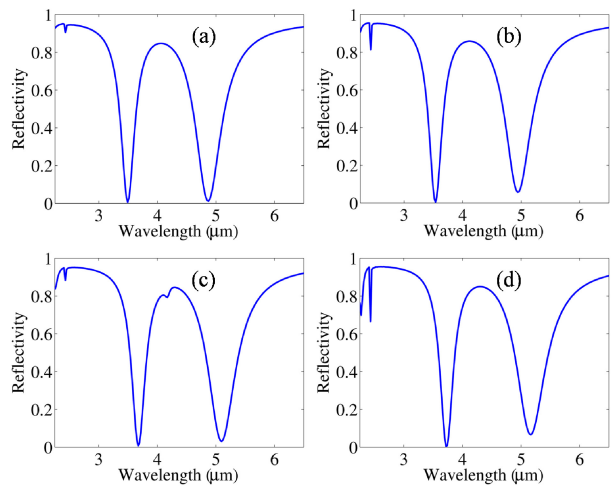

Fig. 2. Calculated optical power reflectivity from dual-spectral band absorbers with: (a) 0.90 μm and 1.30 μm gold square sizes and 77 nm gold film thickness, (b) 0.90 μm and 1.30 μm gold square sizes and 100 nm film thickness, (c) 0.95 μm and 1.35 μm gold square sizes and 77 nm film thickness, (d) 0.95 μm and 1.35 μm gold square sizes and 100 nm gold film thickness.

**C. Design of triple-spectral band metamaterial perfect absorbers**

Triple spectral-band perfect absorbers were also designed and simulated by using the FDTD code. The triple spectral-band perfect absorbers have four thin film gold squares of three different sizes in the unit cell. Once again, the periods of the unit cells are all 2.7 μm in both the x and y dimensions and the thickness of the $MgF_2$ spacer layer is still kept at 75 nm. The optical reflectivity from a triple-band absorber with 0.70 μm, 1.00 μm and 1.38 μm gold squares in the unit cell is shown in Fig. 3 (a) and (b) for gold square thicknesses of 77 nm and 100 nm, respectively. The absorption reaches peak values of 99.7% at 2.88 μm wavelength, 99.5% at 3.85 μm wavelength, and 99.8% at 5.21 μm wavelength for the 77 nm thick structure and 97.0% at 2.84 μm wavelength, 99.6% at 3.89 μm wavelength, and 98.2% at 5.32 μm wavelength for the 100 nm thick structure. In Fig. 3 (c) and (d), the gold squares sizes are increased to 0.75 μm, 1.05 μm and 1.43 μm. The gold film thickness of 77 nm, shown in Fig. 3(c), gives absorption peaks of 99.8% at 3.01 μm, 98.4% at 4.06 μm, and 99.6% at 5.41 μm while the 100 nm film thickness, shown in Fig. 3 (d), results in 99.0% at 3.01 μm, 99.9% at 4.08 μm, and 97.6% at 5.45 μm. It can be seen that the design shown in the Fig. 3(a) has the best performance.

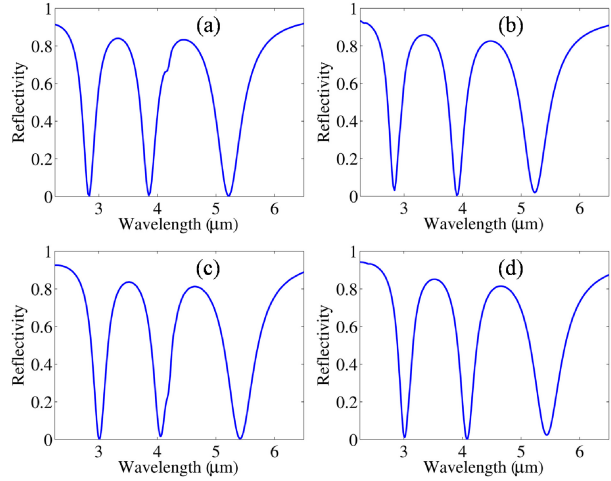

Fig. 3. Calculated optical power reflectivity from triple-band metamaterial absorbers with: (a) 0.70 μm, 1.00 μm and 1.38 μm gold squares of 77 nm gold film thickness, (b) 0.70 μm, 1.00 μm and 1.38 μm gold squares of 100 nm gold film thickness, (c) 0.75 μm, 1.05 μm and 1.43 μm squares of 77 nm gold film thickness, (d) 0.75 μm, 1.05 μm and 1.43 μm gold squares of 100 nm gold film thickness.

## 3. Experimental Results

To fabricate the multi-spectral band metamaterial perfect light absorbers, we first evaporated a 20 nm thick titanium adhesion layer onto a polished silicon wafer. Next, a 250 nm gold film was sputtered on top of the titanium layer followed by a 75 nm thick magnesium fluoride ($MgF_2$) film serving as the dielectric spacing layer. An e-beam resist layer of PMMA 495k was then spin-coated on top of the $MgF_2$ dielectric layer. Using electron beam lithography, multiplexed square openings of different sizes were patterned in the e-beam resist layer which was then developed in a 1:3 MIBK/IPA solution. Following development, an oxygen plasma descum process was conducted to remove the e-beam resist residues in the opened square areas. A 2 nm chromium layer was then evaporated as an adhesion layer for the subsequent gold film evaporation. Finally, a lift-off process was carried out in an acetone solution to remove the e-beam resist and the unwanted gold areas, leaving behind only the gold squares. The total area of each device is 100 μm by 100 μm and the absorber device areas are separated by 3 mm distance from each other. Fig. 4 (a) shows the atomic force microscope (AFM) image of a dual-band metamaterial absorber surface with two gold squares of 0.95 μm and 1.35 μm sizes in the unit cell. The periods of the gold squares are 2.4 μm in both the x and y dimensions. Fig. 4 (b) shows a triple-band metamaterial absorber surface with four gold squares of three different sizes of 1.43 μm, 1.05 μm and 0.75 μm in the unit cell. The periods of the unit cell for the triple band absorbers are 2.7 μm in both the x and y dimensions.

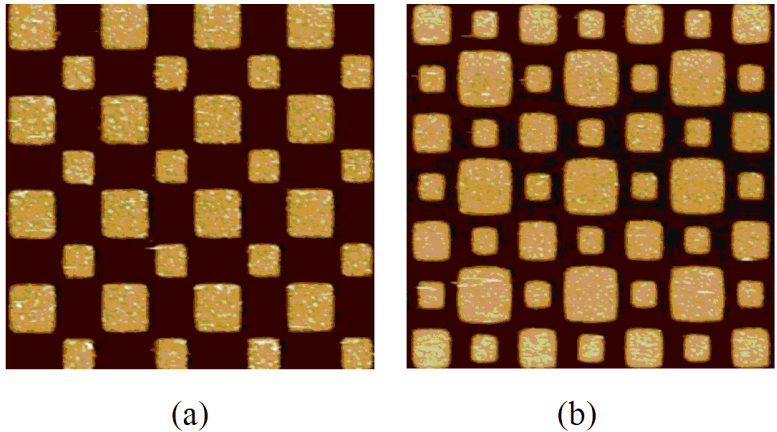

(a)          (b)

Fig. 4. Atomic force microscope (AFM) images of (a) a dual-spectral band metamaterial absorber surface structure with two different size gold squares of 0.95 μm and 1.35 μm in the unit cell and (b) a triple-spectral band metamaterial absorber surface structure with four gold squares of three different sizes of 1.43 μm, 1.05 μm and 0.75 μm multiplexed in the unit cell.

Optical power reflectivity was measured by using a 36x objective microscope coupled Fourier transform infrared spectrometer (Hyperion microscope and Bruker Vertex 80V FTIR) with a mercury-cadmium-telluride (MCT) photodetector. Each spectrum was obtained by averaging 64 measurement scans with 2 cm$^{-1}$ resolution. The measured device area was 50 μm by 50 μm in the center of each device, achieved by appropriately aperturing down the reflected light. Fig. 5 (a) shows the measured optical reflectivity from the dual-band absorber with 0.90 μm and 1.30 μm gold squares of 77 film thickness in the unit cell. The absorption reaches one peak of 90.2% at 3.34 μm wavelength and another peak of 94.7% at 4.63 μm wavelength. Fig. 5 (b) shows the measured optical reflectivity from the dual-band absorber with 0.90 μm and 1.30 μm gold squares of 100 nm film. The absorption now reaches 92.6% at 3.36 μm and 94.7% at 4.67 μm. It is seen from Figs. 5 (a) and (b) that the absorber with 100 nm thick gold film squares gives slightly better performance than the absorber with 77 nm thick gold film. Fig. 5 (c) shows measured optical reflectivity from a two dual-band absorber with 0.95 μm and 1.35 μm squares of 77 nm gold film in the unit cell. The absorption reaches a peak of 92.5% at 3.46 μm and 94.5% at 4.80 μm. Fig. 5 (d) shows measured optical reflectivity from a two dual-band absorber with 0.95 μm and 1.35 μm squares of 100 nm gold film in the unit cell. This dual-band absorber has an absorption peak of 96.0% at 3.50 μm and an absorption peak of 95.0% at 4.82 μm, respectively.

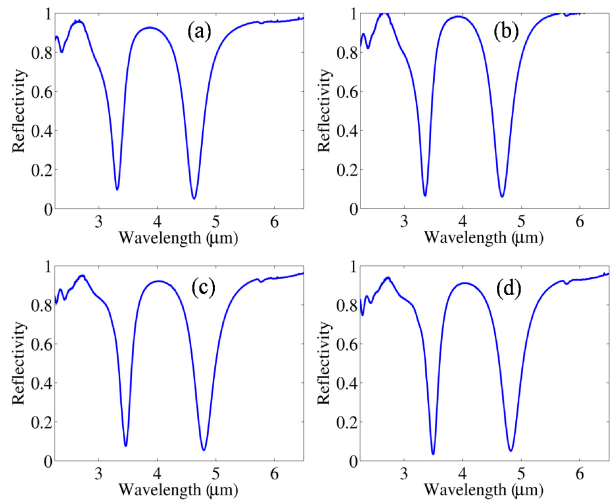

Fig. 5. Measured optical reflectivity from dual-spectral band metamaterial absorbers with: (a) 0.90 μm and 1.30 μm gold squares of 77 nm gold film thickness, (b) 0.90 μm and 1.30 μm gold squares

of 100 nm gold film thickness, (c) 0.95 μm and 1.35 μm gold squares of 77 nm gold film thickness, (d) 0.95 μm and 1.35 μm gold squares of 100 nm gold film thickness.

We also fabricated four triple spectral-band absorbers with two different gold square sizes and of two different film thicknesses. Fig. 6 (a) shows measured optical reflectivity from a triple spectral-band absorber with 0.70 μm, 1.00 μm, and 1.38 μm gold squares in the unit cell. The gold film thickness of the squares is 77 nm. The absorption reaches peaks of 80.9%, 90.3%, and 95.3% at wavelengths of 2.59 μm, 3.66 μm, and 4.95 μm, respectively. In Fig. 6 (b), the film thickness of the gold squares is now 100 nm and the absorption reaches peaks of 80.5% at 2.60 μm, 91.0% at 3.70 μm, and 94.3% at 4.99 μm, respectively. Fig. 6 (c) shows the measured optical reflectivity from a triple-band absorber with 0.75 μm, 1.05 μm and 1.43 μm squares in the unit cell with a thickness 77 nm. The absorption reaches peaks of 93.0% at 2.73 μm, 94.5% at 3.83 μm, and 96.1% at 5.13 μm. Increasing the gold film thickness to 100 nm as shown in Fig. 6 (d), results in absorption peaks of 93% at 2.73 μm, 94.6% at 3.84 μm, and 96.1% at 5.13 μm, respectively.

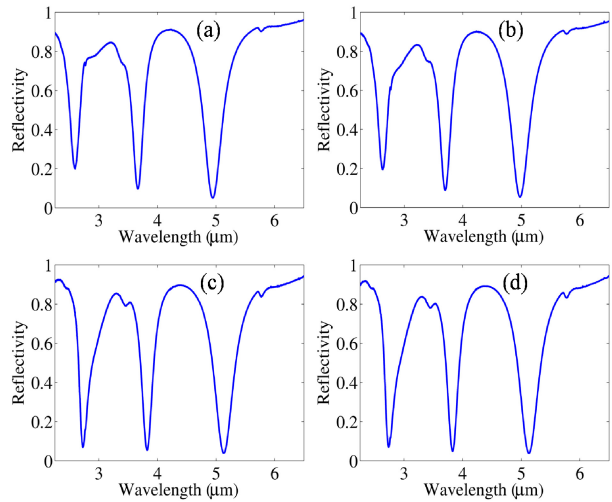

Fig. 6. Measured optical reflectivity from triple-band metamaterial absorbers with: (a) 0.70 μm, 1.00 μm and 1.38 μm gold squares of 77 nm gold film thickness, (b) 0.70 μm, 1.00 μm and 1.38 μm gold squares of 100 nm film thickness, (c) 0.75 μm, 1.05 μm and 1.43 μm squares of 77 nm film thickness, (d) 0.75 μm, 1.05 μm and 1.43 μm gold squares of 100 nm film thickness.

## 4. Discussions

To understand multi-spectral near perfect absorption in multiplexed metal square structures, we simulated the electric field distribution in the middle of the MgF$_2$ spacer layer at the peak absorption wavelengths for an absorber with two thin film gold squares of 0.90 μm and a 1.30 μm in the unit cell. The thickness of the gold film is 77 nm. Two dimensional (2D) distributions of the real and imaginary parts of the x and z components of the electric field ($E_x$ and $E_z$) are plotted in Fig. 7 at the 3.49 μm peak absorption wavelength and plotted in Fig. 8 at the 4.87 μm peak absorption wavelength. It can be seen that the electric field is enhanced one order of magnitude underneath the 0.9 μm gold square at the peak absorption wavelength of 3.49 μm and also enhanced one order of magnitude underneath the 1.3 μm gold square at the peak absorption wavelength of 4.87 μm.

From the electric field distributions shown in Figs. 7 and 8, it can be seen that the z-component of the electric field ($E_z$) is significantly enhanced at the two ends of the excited gold squares at the respective peak absorption wavelengths. The $E_z$ component at the center location below the metal square is zero at the peak absorption wavelengths and oscillates out of phase at the two ends. The x-component of the electric field ($E_x$) is also significantly enhanced at the two ends of the excited gold squares at the peak absorption wavelengths, however, the $E_x$ component oscillates in phase at the two ends. The $E_x$ component at the center location below the metal square is very small, less than 0.05 V/m. The electric field distribution indicates that a quadrupole mode resonance is excited at the resonance wavelength corresponding to each metal square in the structure. The quadrupole resonance is formed by the dipole resonance of the metal square on the top of the dielectric spacer layer and the induced image dipole resonance in the thick gold layer. Two electric dipoles oscillate out of phase and form a quadrupole. The quadrupole resonance is essentially the same as the magnetic resonance induced by the anti-parallel electrical currents in the THz regime [41-43]. Because the dielectric spacer layer thickness is only a small fraction of the wavelength, the absorption in the absorber is primarily due to the local quadrupole mode plasmon resonance. The small x component of the electric field in the center location ($E_x$) is due to the non-local guided mode propagating in the x-y plane. The guided mode is excited by the periodic gold square array. The electric field distributions show that each excited metal square in the metamaterial structure gives rise to the near perfect light absorption at the corresponding quadrupole mode plasmon resonance wavelength.

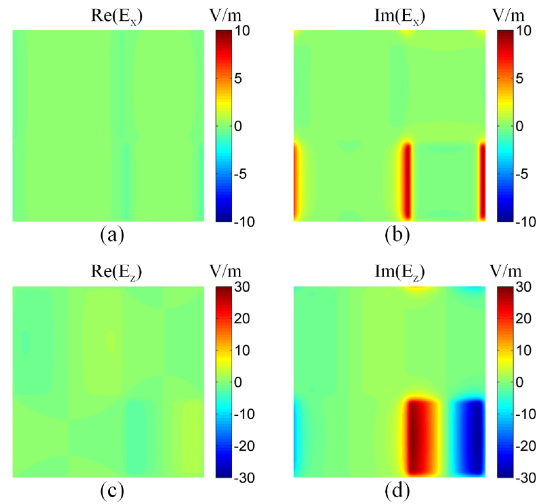

Fig. 7. Simulated electric field distributions in the middle of the dielectric spacer layer of a dual-band perfect absorber at 3.49 μm wavelength: (a) the real part of the x-component of the electric field, (b) the imaginary part of the x-component of the electric field, (c) the real part of the z-component of the electric field, (d) the imaginary part of the z-component of the electric field.

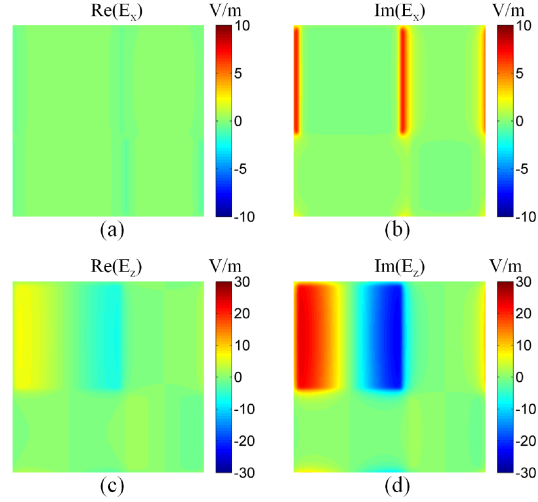

Fig. 8. Calculated electric field distributions in the middle of the dielectric spacer layer of a dual-band perfect absorber at 4.87 μm wavelength: (a) the real part of the x-component of the electric field, (b) the imaginary part of the x-component of the electric field, (c) the real part of the z-component of the electric field, (d) the imaginary part of the z-component of the electric field.

The electric field profiles in the middle of the $MgF_2$ dielectric spacer layer at the three near peak absorption wavelengths of a triple-band absorber were also simulated. The studied structure has 0.7 μm, 1.0 μm and 1.38 μm gold squares in the unit cell of 77 nm thickness. The distributions of the real and imaginary parts of the x and z components of the electric field at the peak absorption wavelength of 2.88 μm are plotted in Figs. 9 (a-c), respectively. It can be seen that both the x and z components of the electric field are strongly enhanced underneath the 0.7 μm gold squares (top right and bottom left). We also simulated the electric field distributions in the middle of the dielectric spacer layer at the other two peak absorption wavelengths; 3.85 μm and 5.21 μm. As anticipated, the 1.0 μm gold square and the 1.38 μm square are excited at their corresponding peak absorption wavelengths. The electric field distributions in the triple-band absorber metamaterials further confirm that the quadruple mode plasmonic resonances in the multiplexed gold metal squares contribute to the near perfect light absorption at the resonance wavelengths.

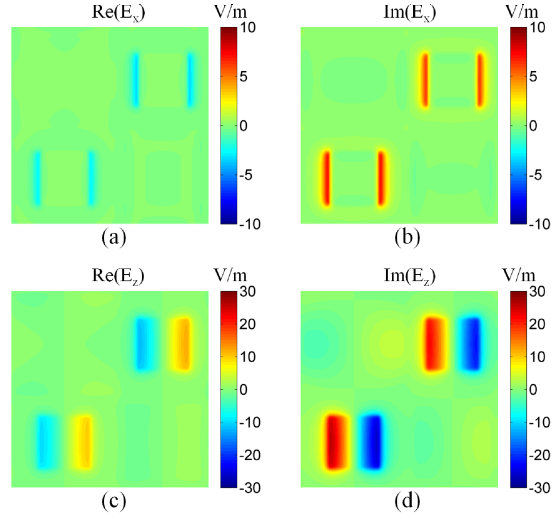

Fig. 9. Simulated electric field distributions in the middle of the dielectric spacer layer of a triple-band perfect absorber at 2.88 μm wavelength: (a) the real part of $E_x$ component, (b) the imaginary part of $E_x$ component, (c) the real part of $E_z$ component, (d) the imaginary part of $E_z$ component.

## 5. Summary

In summary, we designed and experimentally demonstrated several dual spectral-band and triple spectral-band near perfect metamaterial absorbers in the mid-wave IR regime using spatially multiplexed metal square structures. For the dual-band absorbers, optical power absorptions over 95% at two separate wavelengths were observed. For the triple-band metamaterial absorbers, optical absorptions over 92.5% were observed at three separate wavelengths. The experimental results show that the peak absorption wavelengths of the metamaterial absorbers are primarily controlled by the sizes of the metal squares. We simulated the electrical field distributions in the middle plane of the dielectric spacing layer at the peak absorption wavelengths. The electric field distributions show that the strong light absorptions correspond to the local quadrupole mode plasmon resonances in the multiplexed metal square metamaterial structures.

## Acknowledgments

B. Zhang and J. Guo acknowledge the support by the National Science Foundation (NSF) through the grant NSF-0814103. J. Hendrickson would like to acknowledge support from the Air Force Office of Scientific Research (Program Manager Dr. Gernot Pomrenke) under contract number 12RY05COR. The authors would also like to thank Jodie Shoaf for AFM measurements.